# Low-Power consumption Franz-Keldysh effect plasmonic modulator

N. Abadía,[1,2,*] T. Bernadin,[3] P. Chaisakul,[2] S. Olivier,[1] D. Marris-Morini,[2] R. Espiau de Lamaëstre,[1] J. C. Weeber,[3] and L. Vivien[1]

[1]*CEA, LETI, MINATEC Campus, 17 rue des Martyrs, F-38054 Grenoble, France*
[2]*Institut d'Electronique Fondamentale, Université Paris-Sud 11, CNRS UMR 8622, Bât. 220, F-91405 Orsay, France*
[3]*Laboratoire de Physique, Université de Bourgogne, CNRS UMR 5027, 9 Avenue A. Savary, F-21078 Dijon, France*
*[*]nicolas.abadia@u-psud.fr*

**Abstract:** In this paper we report on a low energy consumption CMOS-compatible plasmonic modulator based on Franz-Keldysh effect in germanium on silicon. We performed integrated electro-optical simulations in order to optimize the main characteristics of the modulator. A 3.3 dB extinction ratio for a 30 µm long modulator is demonstrated under 3 V bias voltage at an operation wavelength of 1647 nm. The estimated energy consumption is as low as 20 fJ/bit.

©2014 Optical Society of America

## 1. Introduction

The electronic industry tries to improve the transistor performance by downscaling the dimensions of the devices. Nowadays it is believed that the Moore's law, which states that the density of transistor in the surface of an integrated circuit doubles every two years for a constant cost, has reached its limit. When the transistors are downscaled, the dimensions of the electrical lines that interconnect them are also reduced. The parasitic capacitance between adjacent wires is also increased. This causes an increase in the propagation time delay of the electrical signal. This problem is known as the interconnect bottleneck. To overcome it, it has been proposed to substitute electrical links by optical ones. Silicon photonics technology is one of the options, whose advantage is the integrated fabrication of photonic integrated circuits (PICs) at a silicon wafer scale and the use of existing microelectronic processes.

Several devices based on silicon photonics have been proposed such as modulators [1], detectors [2] and sources [3]. The photonic modulator plays an important role towards the fabrication of photonics integrated chips [4]. Using the free carrier dispersion effect in silicon (Si) several electro-refractive modulators have been proposed. They use a Mach-Zehnder Interferometer (MZI) or a ring resonator structure to convert the phase shift brought by the change of refractive index into a modulation of the input power. A drawback of the MZI structure is that a millimeter long device is required to achieve a significant modulation, leading to high energy consumption in the order of pJ/bit. The power consumption can be reduced by using a ring resonator structure but it is sensitive to fabrication defects, due to its highly resonant character. On the other hand, electro-absorption devices provide shorter structures with low energy consumption and high operational frequencies. Two physical effects can be used in electro-absorption modulators: the Franz-Keldysh effect (FKE) [5,6] in bulk germanium (Ge) and the Quantum Confined Stark Effect (QCSE) [7] in Ge/SiGe quantum wells.

To further reduce the photonic modulator footprint and thus power dissipation, use of plasmonic modes has been proposed. Indeed a higher level of optical confinement, hence of electro-optical interaction, can be expected. Several plasmonic modulators based on carrier dispersion effects have been proposed, e.g. using a MZI between a photonic mode and a plasmonic one [8], or between two plasmonic modes [9,11]. Electro-absorption plasmonic modulators in Si [12], Indium tin Oxide (ITO) [10], using a stub with QCSE [13] and a dielectric gain medium $Er^{3+}$ [14] have also demonstrated promising performance.

In this paper, we propose a Ge FKE plasmonic electro-absorption modulator compatible with CMOS fabrication technology. Such modulator is based on a vertical MIS (Metal-Insulator-Semiconductor) waveguide composed of copper (Cu), silicon nitride (Si3N4) and a non-intentionally doped (nid) tensile strained Ge core on Si.

## 2. Structure of the device

The structure consists of a vertical MIS waveguide formed by Cu, $Si_3N_4$ and a nid Ge core. The cross-section of the structure is represented in Fig. 1. Most of published plasmonic devices used either silver (Ag) [8] or gold (Au) [14] as a metal. However these metals are prohibited in CMOS environment since they are considered as contaminants. Conversely, other metals like Cu and aluminum (Al) are CMOS compatible. In our structure we select Cu due to the lower optical losses with respect to Al [15]. We use a low optical loss Cu. The complex refractive index was measured by ellipsometry and is given by $n_{Cu}$ = 0.657 + 9.61i at a wavelength of 1647 nm. The $Si_3N_4$ slot has a thickness $h_{Slot}$ and the nid Ge core has a width w and a height h. We use $Si_3N_4$ as both the insulator of the capacitance and the diffusion barrier of Cu into the Ge core. It was shown experimentally that it allows a reliable fabrication of integrated plasmonic devices in a CMOS environment [15]. Furthermore, the presence of the $Si_3N_4$ slot reduces the optical losses of the MS (Metal-Semiconductor) plasmonic waveguide mode since it reduces the penetration depth of the optical electric field in the metal. Such a MIS structure is disposed over a p-doped region of Ge and a p-doped region of Si. One contact of the structure is taken in the Cu metal while the other is taken in the p-doped Ge and Si layers at the bottom of the structure. Such contacts are used to apply a differential voltage V and induce a static electric field in the Ge core waveguide. The side contact at the bottom is connected to ground while the top contact is set to the desired voltage.

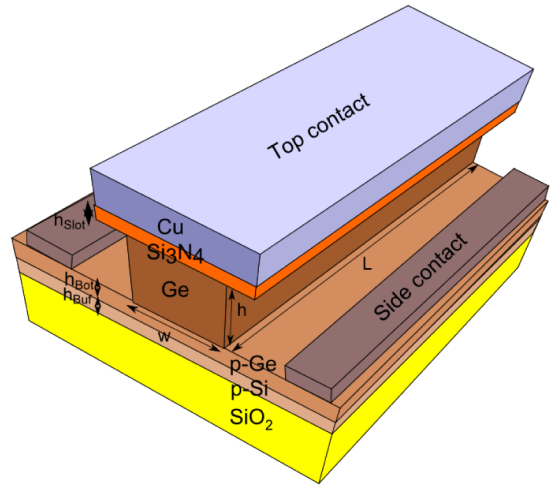

Fig. 1. Structure of the Franz-Keldysh effect plasmonic modulator. The modulator is surrounded by $SiO_2$.

## 3. Models

In order to perform an integrated optoelectronic simulation of our plasmonic modulator device, we simulate the structure electrically using the commercial software ISE-DESSIS. After solving the static electric field distribution in the structure, we are able to calculate the changes in the losses of the material $\Delta\alpha_{mat}$ in the Ge part due to the FKE. Using the induced FKE loss distribution, we subsequently simulate the structure optically by using a finite difference method (FDM) mode solver. From these simulations, we obtain the change in the effective losses $\Delta\alpha_{eff}$ of the plasmonic mode as a function of the applied voltage V in the structure, so that the voltage dependence of the extinction ratio (ER) as well as the propagation losses (PL) of the modulator can be calculated.

The generalized FKE formalism described in [16] is used to simulate the Ge absorption variation under an electrical field:

$$\Delta\varepsilon(E,F) = \left(\frac{B}{E^2}\right)(\bar{h}\theta)^{\frac{1}{2}}\left[G(\eta) + iF(\eta)\right] \quad (1)$$

$$\bar{h}\theta = \left(\frac{e^2 \bar{h}^2 F^2}{2\mu}\right)^{1/3} \quad (2)$$

$$\eta = \frac{E_g - E}{\hbar\theta} \quad (3)$$

where $\Delta\varepsilon$ is the change in the dielectric constant of Ge for photon energy E under an applied electric field F. The parameter B quantifies the probability of photon absorption. $G(\eta)$ and $F(\eta)$ [16] are formulas containing Airy functions. The elementary charge is denoted as e, h is the Planck constant, μ is the reduced effective mass of carriers and $E_g$ is the energy bandgap of the semiconductor. Operating over equation Eq. (1) we can calculate the change in the absorption of the material $\Delta\alpha_{mat}$.

The values for Ge of the parameters B and μ of equations Eq. (1) and Eq. (2) were taken from [18]. In the structure reported in Fig. 1, the Ge is deposited on Si. Consequently the Ge is strained, leading to an enhanced FKE [18]. The operation wavelength at which the FKE is maximum is around 1647 nm [18]. This wavelength will therefore be chosen for modulator operation. According to reference [18] the maximum change of the real part of the refractive index of Ge due to the presence of an electric field is around $8 \times 10^{-4}$ at 1647 nm for an electric field of 60 kV/cm.

When a voltage is applied to the structure of Fig. 1 there is a carrier accumulation in the interface between the $Si_3N_4$ and the Ge core. This changes the real part of the refractive index of Ge and maybe it is in the same order of magnitude that the one obtained in Si.

## 4. Simulation results

The fundamental plasmonic mode of the MIS structure features an electric field distribution partly confined within the thin insulator layer $Si_3N_4$ and delocalized in the Ge core. The normalized optical intensity distribution of such mode is represented in Fig. 2. For the particular case of w = 150 nm, h = 250 nm, $h_{Slot}$ = 5 nm, $h_{Buf}$ = 60 nm and $h_{Bot}$ = 40 nm, we found that 62% of the plasmonic optical mode intensity is in the Ge core while 11% of it is in the $Si_3N_4$ slot. When a bias voltage V is applied between both contacts, a static electric field appears in the Ge core. The distribution of the static electric field in the structure for V = 0 V and V = 3 V are represented in Fig. 3. At 0V, there is no electric field in the Ge core. The static electric field in the $Si_3N_4$ slot is due to the residual charges present in the intrinsic Ge where the residual charge density is $10^{16}$ cm$^{-3}$, and to the presence of the Cu metal on the other side. At 3V, a static electric field appears in the Ge core and changes the losses of the material $\alpha_{mat}$, which in turn leads to the modulation of the plasmonic mode by changing its effective damping constant (or loss constant) $\alpha_{eff}$.

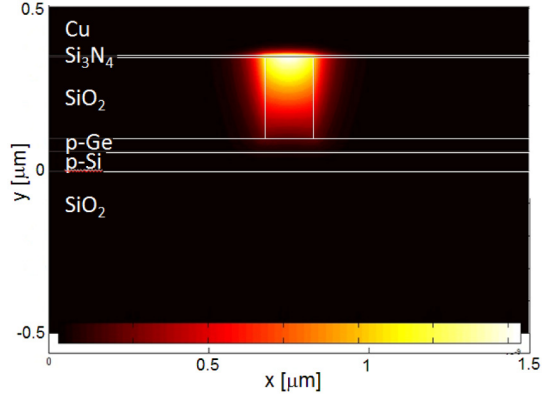

Fig. 2. Normalized intensity distribution of the plasmonic mode supported by the vertical MIS structure. The parameters of the structure are the followings: w = 150 nm, h = 250 nm, $h_{Slot}$ = 5 nm, $h_{Buf}$ = 60 nm and $h_{Bot}$ = 40 nm.

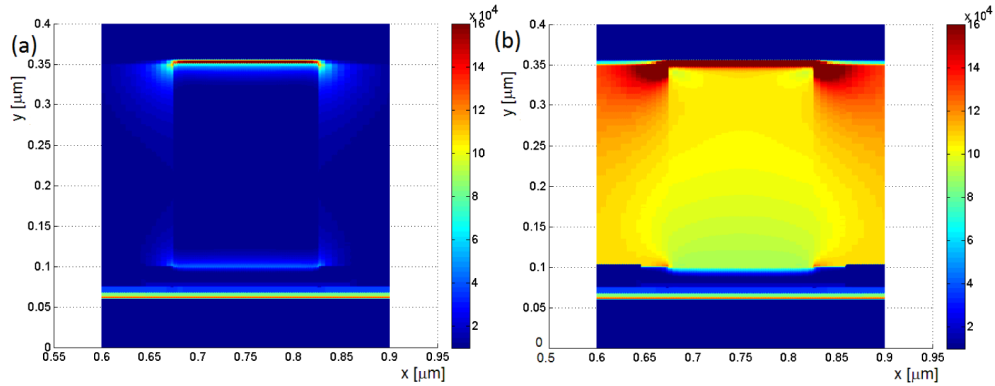

Fig. 3. Distribution of the static electric field in V/cm for an applied voltage V = 0 V (a) and V = 3 V (b). The parameters of the structure are the followings: w = 150 nm, h = 250 nm, $h_{Slot}$ = 5 nm, $h_{Buf}$ = 60 nm and $h_{Bot}$ = 40 nm.

We optimized the structure with respect to w and h. $h_{Slot}$ has to be below 10 nm in order to reduce the fraction of the static electric field in the $Si_3N_4$ slot. When the $Si_3N_4$ barrier is thinner, the static electric field distribution is more deconfined from the $Si_3N_4$ slot into the Ge core. Consequently there is more static electric field in the Ge core where the FKE is present. However, $h_{Slot}$ has to be thicker than 3 nm to efficiently act as a diffusion barrier between the Cu metal and the Ge core. A value of $h_{Slot}$ = 5 nm is finally selected to avoid working in the 3 nm limit and account for fabrication tolerances. The parameters w and h have a direct influence on the figure of merit $\Delta\alpha_{eff}/\alpha_{eff}$ of the modulator where $\Delta\alpha_{eff}$ is the change in $\alpha_{eff}$ due to the FKE in the Ge core and it is equal to $\Delta\alpha_{eff} = \alpha_{eff}$(V = 3 V, ON state)-$\alpha_{eff}$(V = 0 V, OFF state). This figure of merit is useful to optimize the trade-off between the ER and the PL of the modulator. The optimization of the parameter $\Delta\alpha_{eff}/\alpha_{eff}$ with respect to w and h for different bias voltages is presented in Figs. 4(a) and 4(b).

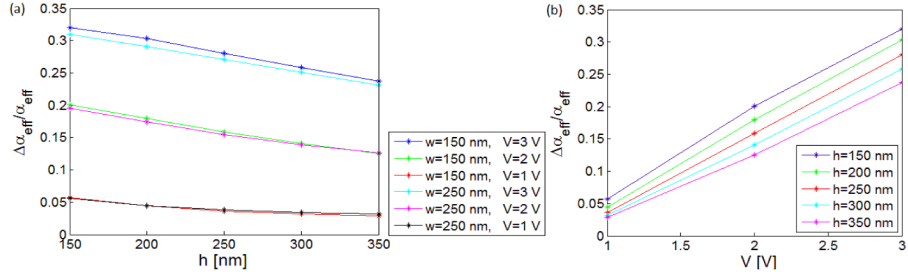

Fig. 4. (a) Optimization of the FoM $\Delta\alpha_{eff}/\alpha_{eff}$ as a function of w and h for different driving voltages V. Two cases are selected: w = 150 nm and w = 250 nm, (b) Optimization of the parameter $\Delta\alpha_{eff}/\alpha_{eff}$ as a function of the driving voltage V for w = 150 nm.

Figure 4(a) presents $\Delta\alpha_{eff}/\alpha_{eff}$ as a function of the Ge height h for different modulator waveguide widths w. We can conclude that the thinner the width of the Ge core w, the higher $\Delta\alpha_{eff}/\alpha_{eff}$. We selected w = 150 nm since it is the minimum width for which a TM plasmonic mode can be supported and this dimension is roughly the minimum feature that can be fabricated using 193 nm deep-UV lithography. From Fig. 4(a), it can also be concluded that the thinner the parameter h, the better. Furthermore, as expected, the efficiency $\Delta\alpha_{eff}/\alpha_{eff}$ is increasing with the driving voltage V (Fig. 4(b)). To limit the power consumption of the device we decided to work at a maximum voltage of V = 3 V. Setting V, the thickness h of the Ge core must be chosen in such a way that the maximum static electric field in the Ge core is below the breakdown voltage of Germanium ($10^5$ V/cm). According to our calculations, for V = 3 V, the minimum value of h is 250 nm compatible with the epitaxial growth of Ge on Si. Note that with this configuration, the static electric field within the insulator barrier is also well below the breakdown voltage of $Si_3N_4$. Furthermore, such a value allowed the epitaxial growth of Ge on Si including buffer layer. For those values, an almost uniform static electric field (see Fig. 3) close to $10^5$ V/cm is present in the Ge core.

For the optimized value w = 150 nm, the computed ER and the PL of the modulators are plotted in Fig. 5(a) and 5(b) respectively as a function of h. For the targeted value of h = 250nm, an ER of 3.3 dB with PL of 11.2 dB can be achieved for a device length of L = 30 µm. If the device length L is increased, then the ER is also increased at the cost of larger PL. The ER per unit length is 0.11 dB/µm and the PL is 0.37 dB/µm. To calculate the coupling loss (CL) of the device preliminary results shows that it is around 1 dB using a butt-coupling scheme between a rib waveguide and the modulator. The operation frequency is RC limited and it is expected to be around hundreds of GHz. Using $f_c = 1/2\pi RC$, with R = 50 Ω the resistance and C = 9 fF the capacitance of the device of length 30 µm simulated by ISE-DESSIS, we find a cutoff frequency of about 350 GHz. For the calculation of the equivalent resistance, the contact resistance of the side contact was taken into account (See Fig. 1) as well as the series resistances of the p-Ge and the p-Si layers for a distance between the side contact and the Ge core of 1.5 µm. The doping density of the p-Ge layer and p-Si layers are $10^{19}$ cm$^{-3}$. The height of the p-Ge layer and p-Si layers are 40 and 60 nm respectively.

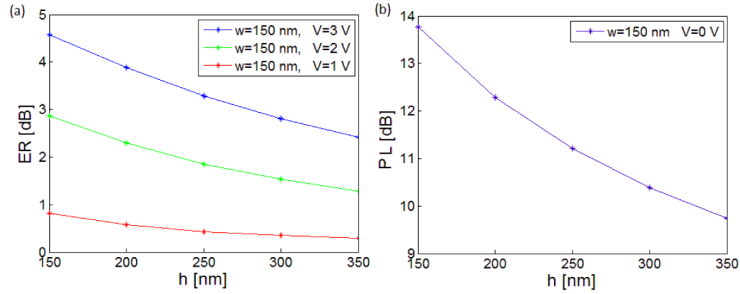

Fig. 5. (a) Extinction ratio of the plamonic modulator as a function of h for different driving voltages V and w = 150 nm. The longitude of the plasmonic modulator is L = 30 μm, (b) PL of the plasmonic modulator for a driving voltage V = 0 V and w = 150 nm. The longitude of the plasmonic modulator is L = 30 μm.

Regarding the energy consumption (EC), the model described in [17] is considered. In this model the EC of the dynamic modulation is given by the formula $E_{bit} = 1/4 CV_{DD}^2$ where C is the capacitance of the device and $V_{DD}$ is the driving voltage. Using the previously calculated value of the capacitance C = 9 fF, we find an energy per bit of about 20 fJ/bit. This value is smaller than the previously reported photonic FKE modulators [11,12,15] and plasmonic ones [24] which state EC, listed in Table 1.

**Table 1. Electrical Power Consumption of the FK Effect Modulators**

| Team: | MIT [20] (2008) | A*STAR [12] (2011) | Kotura Inc [22]. (2011) | Leeds (2011) [24] | Berkeley (2012) [25] | Stanford (2013) [26] | Our work (2013) |
|---|---|---|---|---|---|---|---|
| EC [fJ/bit] | 50 | 200 | 100 | 360 | 56 | 100 | 20 |

In order to evaluate the operating wavelength range of the modulator, electro-optical simulations were performed at different wavelengths for the optimal modulator (w = 150 nm, h = 250 nm). The resulting $|\Delta\alpha_{eff}|/\alpha_{eff}$ figure of merit is plotted in Fig. 6(a). The dispersion of the material was taken into account.

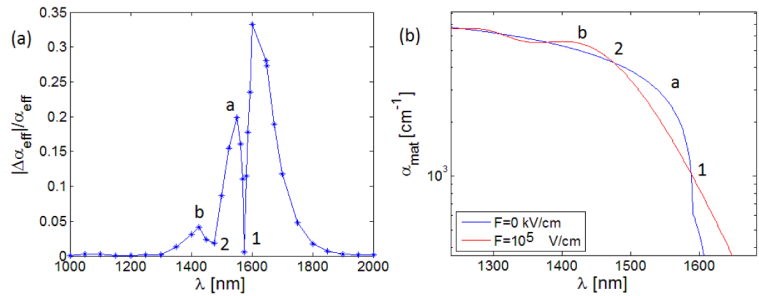

Fig. 6. (a) Simulated figure of merit $|\Delta\alpha_{eff}|/\alpha_{eff}$ for different operating wavelengths of the device. The optimized device: w = 150 nm, h = 250 nm is considered; (b) Material absorption of the Ge due to the FKE for different operational wavelengths.

As expected, the best operational wavelength of the device is around 1647 nm where the FKE is maximum, as we stated before. Above 1647 nm performances are degraded. Franz-Keldysh Oscillations (FKO) occurs below 1550 nm, the direct bandgap of Ge. To explain these oscillations, we plotted the absorption of Ge against the operational wavelength in Fig. 6(b) for zero static electric field and for $10^5$ V/cm, the field corresponding to 3 V applied voltage. At the points marked 1 and 2 in Fig. 6(b), absorption values of the Ge core for 0 V (0 V/cm) and 3 V ($10^5$ V/cm) are equal. Under such condition the parameter $|\Delta\alpha_{eff}|$ is close to zero, hence the drops in the modulator figure of merit in Fig. 6(a) at the points also marked 1 and 2. The maximums a and b of Fig. 6(a) also correspond with the points a and b in Fig. 6(b) where the absorption difference between 0 and 3 V is maximum. As can be seen in Fig. 6(a), the FoM $\Delta\alpha_{eff}/\alpha_{eff}$ reaches a maximum value of 0.34. It is below the FoM of 2-3 reported in some photonic electro-absorption modulators [22,25]. The reason to obtain a lower FoM in the plasmonic modulator is due to the higher effective losses $\alpha_{eff}$, due to the use of metal. However, the FoM of our modulator could be further improved in the future by using nanostructuration. Besides, a plasmonic switch was proposed in [26] with a high FoM of 12, using Vanadium (IV) oxide ($VO_2$). However the operational frequency is in the kHz regime since it is based on thermal modulation.

The useful wavelength operating range is 1590 to 1650 nm. Promisingly, an inclusion of small amount of Si (0.75%) into Ge was shown to enable modulation at the telecommunication wavelength of 1.55 µm [20].

## 5. Conclusion

In this paper, we proposed a new design of a low energy consumption germanium FKE electro-absorption plasmonic modulator. We have performed an integrated electro-optical simulation in order to determine the performances of the device. Using the static electric field distribution we calculated the change in absorption $\Delta\alpha_{mat}$ occurring in germanium due to the Franz-Keldysh effect. Finite difference method optical simulations were performed in order to determine the extinction ratio and the insertion losses of the modulator. Using this method, we optimized the structure in order to maximize the figure of merit $\Delta\alpha_{eff}/\alpha_{eff}$. The optimized device has a compact active region of w = 150 nm and h = 250 nm and is fully compatible with the CMOS fabrication technology. For a length of L = 30 µm the device has an extinction ratio of 3.3 dB and an insertion losses of 11.2 dB, while working at a bias voltage of V = 3 V. High-speed operation can be achieved with a cut-off frequency beyond 300 GHz. The energy consumption of such a device is around 20 fJ/bit which is well below the energy consumption of previous reported modulators [18–23].

**Acknowledgment**

This research is part of the MASSTOR project supported by ANR (ANR-11-NANO-022).